\documentclass[useAMS,usenatbib,usegraphicx]{mn2e}

\usepackage{amsmath}
\usepackage{amsfonts}
\usepackage{rotating}
\usepackage{color}
\usepackage{pdflscape}
\usepackage{txfonts}
\usepackage{natbib, twoopt}
\usepackage[breaklinks=true]{hyperref} 
\bibpunct{(}{)}{;}{a}{}{,} 

\title[]{NGC~4337: an over-looked old cluster in the inner disc of the Milky Way \thanks{Based on observations obtained at the Cerro Tololo Inter-American Observatory, Chile }} 

\author[]{Giovanni Carraro$^{1}$\thanks{On leave of absence from Padova University. E-mail: gcarraro@eso.org}, Edgard E. Giorgi$^{2}$, Edgardo Costa$^{3}$, Ruben A. V\'azquez$^{2}$\\
$^{1}$ESO, Alonso de Cordova 3107, 19001, Santiago de Chile, Chile\\
$^{2}$ Facultad de Ciencias Astron\'omicas y Geof<ED>sicas (UNLP), Instituto de Astrof\'isica 
de La Plata (CONICET, UNLP), Paseo del Bosque s/n, La Plata, Argentina\\
$^{3}$ Departamento de Astr\'onomia, Universidad de Chile, Casilla 36-D, Santiago de Chile, Chile}
\begin{document}



\maketitle

\label{firstpage}

\begin{abstract}
Galactic open clusters do not survive long in the high density regions of the inner Galactic disc.  Inside the solar ring only  11 open clusters are known with ages  older than one Gyr. We show here, basing on deep, high-quality photometry, that NGC~4337, contrary to earlier findings, is indeed an old open cluster. The cluster is located very close to the conspicuous star cluster Trumpler~20,
as well mis-classified in the past, and that has  received so much attention in recent years. NGC~4337 shows a significant clump of He-burning stars which was not detected previously. Its beautiful 
color-magnitude diagram is strikingly similar to the one of the classical old open clusters IC~4651, NGC~752, and NGC~3680, and this suggests
similar age and composition.
A spectroscopic study is much needed to confirm our findings. This, in turn, would also allow us  to better define the inner disc radial abundance gradient  and  its temporal evolution.To this aim,  a list of clump star candidates is provided.
\end{abstract}
 
\begin{keywords}
(Galaxy): open clusters and associations: general -- (Galaxy): open clusters and associations: individual: NGC 4337
\end{keywords}

\section[]{Introduction}
Recent studies have consolidated a picture of a dual  Milky Way disc. The co-rotation radius roughly 
divides the disc in two regions, the anti-center and the inner regions, that appear to have experienced very different chemical enrichment histories. Among the others, Magrini et al. (2010) and Frinchaboy et al. (2013)  provided plenty of new data that lend strong support to the earlier findings by Twarog et al. (1997) that the radial abundance gradient is not monotonic over the whole disc extent. In agreement with observations in external spiral galaxies (Bresolin  et al. 2012), at about co-rotation, the radial abundance gradient flattens out in the anti-center region, while being quite steep in the inner regions. Most of these evidences come from studies
of Galactic open clusters, especially the oldest ones (Friel 1995). These clusters are of invaluable importance since they allow us
to trace the chemical properties of the disc since its very early assembly. Ideally, by using these clusters, one would be able to understand, e.g.,  how the abundance gradient built up and changed with time. \\

\noindent
However, as widely known, they are very few in number,
and mostly located in the disc periphery, where star clusters can survive longer. In the inner disc only 11 old open clusters are known, whose ages are larger than 1.0 Gyr (see Table~1).
We caution the reader that this number is much smaller than what one can blindly extract from public catalogs, like for instance the WEBDA database\footnote{http://www.univie.ac.at/webda}.
We restricted ourselves  to clusters with a solid determination of the age, located between l=0$^o$ and l=85$^o$  in the first quadrant, and from l=275$^o$  and l=360$^o$  in the fourth quadrant,
and within a Galacto-centric distance smaller than the Sun value (8.5 kpc).
For most clusters listed on WEBDA as older than 1 Gyr 
either the available data or the analysis  are poor (see the discussion, e.g. in Paunzen et al. 2010),  and their age cannot be considered reliable enough.\\
This table contains also NGC~6791, one of the oldest and more metal rich old open clusters in the Milky Way. Its membership to the disc, however,  has been questioned, since overall
its parameters bear more similarity  to the stellar populations in the Galactic bulge (Carraro 2013).\\

\noindent
The most recent member of this restricted family is Trumpler~20.  Originally mis-classified as a young cluster,  
Trumpler~20 was then correctly found to be a rare example of a rich, old open cluster
in the inner parts of the Milky Way disc ( Platais et al. 2008, Seleznev et al.  2010).  This  triggered many follow-up studies (Donati et al. 2014, Carraro et al. 2014, and references therein).\\

\begin{figure*}
\includegraphics[width=0.45\hsize]{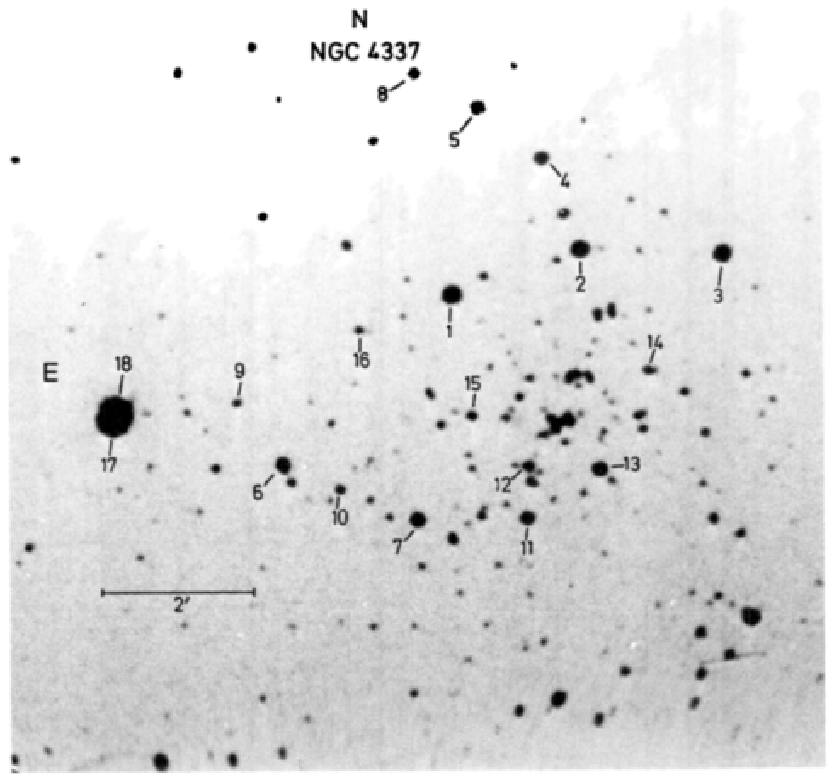}
\includegraphics[width=0.45\hsize]{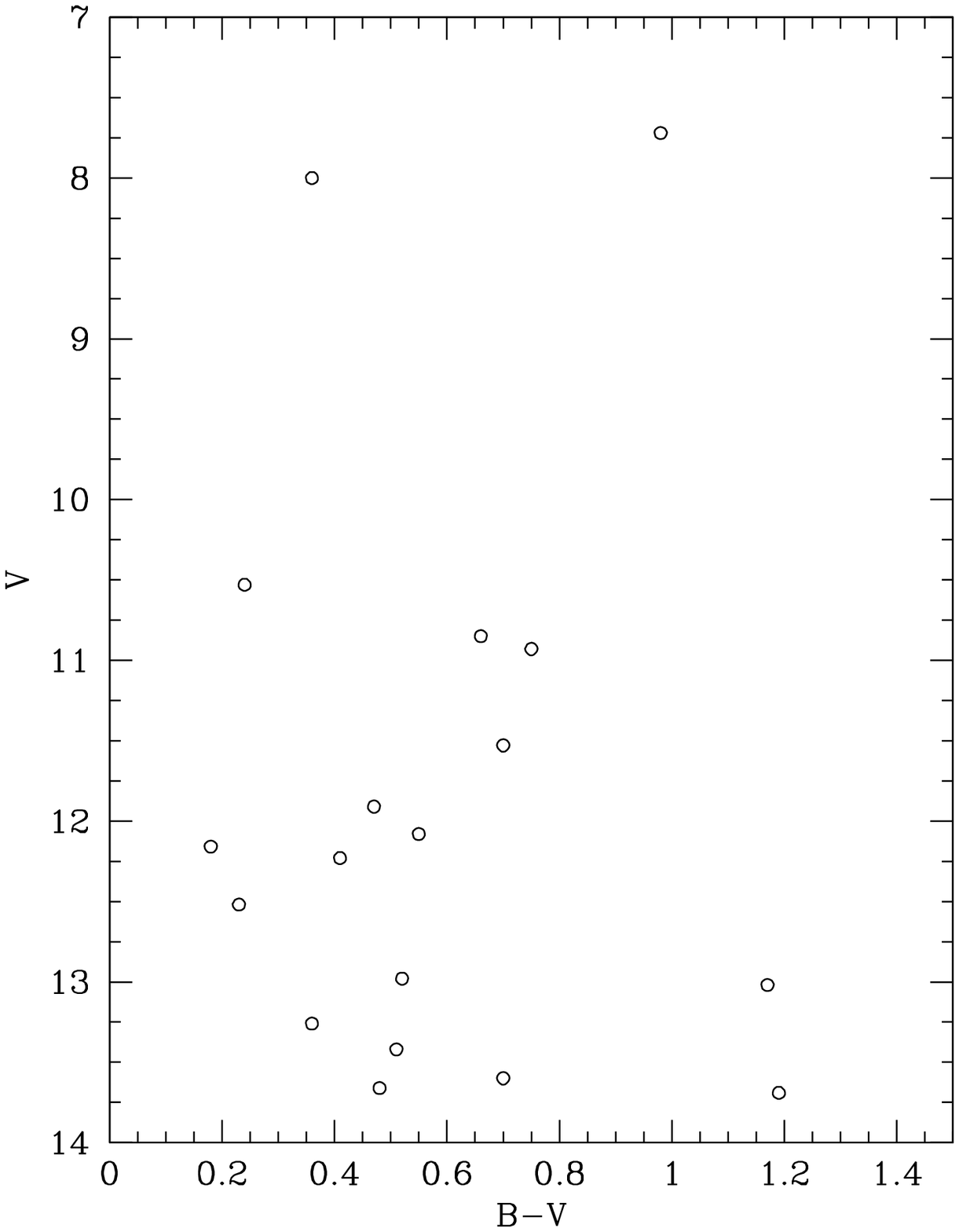}
\caption{{\bf Leff panel:} Image of NGC~4337 from Moffat \& Vogt(1973). Numbers refer to observed stars, as described in the text.
{\bf Right panel:} The CMD derived from Moffat \& Vogt (1973) photometry.}
\end{figure*}

\begin{table}
\tabcolsep 0.1truecm
\caption{Star clusters in the inner disk older than 1.0 Gyr}
\begin{tabular}{lcccccc}
\hline
Cluster & $l$ & $b$ & Age & [Fe/H] & $d_{GC}$ & reference\\
\hline
   & [deg]  & [deg] & Gyr & & kpc & \\
\hline\hline
NGC~6583      &     9.28   &  -2.53    & ~1.0  & +0.37 & 6.5  & Carraro et al. 2005\\
Berkeley~44    &   53.21  &  +3.35   & 1.3                &             &  7.6 &Carraro et al. 2006a\\
Berkeley~52    &   67.89  &    -3.13  & 2.0                &             &  8.1 &Carraro et al. 2006a\\
NGC~6791      &   69.96  & +10.90  & 8.0                & +0.40  & 8.0 & Carraro et al. 2006b\\
NGC~6819      &   73.98  &  +8.48  & 3.0                 &+0.09 &   8.2 & Carraro et al. 2013\\
NGC~3680      & 286.77  & +16.92 & 1.85                & -0.03   &  8.3 & Anthony-Twarog \& Twarog 2004\\
Trumpler~20    & 301.47 & +2.22  & 1.5                 & +0.09 & 7.3  & Carraro et al. 2014\\
Collinder~261 & 301.68 & -5.53   & 7.0                & +0.13 & 7.5  & Gozzoli et al. 2006\\
NGC~6005      & 325.78 & -2.98   & 1.2                &             &  6.5  & Piatti et al 1998\\
NGC~6253      & 335.46 & -6.25   & 3.0                & +0.40 & 7.0 &  Piatti et al. 1998\\
IC~4651           & 340.01 & -7.91   & 1.7               & +0.11 & 8.0 &  Meibom 2000\\
\hline\hline
\end{tabular}
\end{table}

\noindent
We report  here on another very similar case: the old open cluster  NGC~4337.\\
This cluster has 2000. equinox equatorial coordinates RA= $12^{h}24^{m}04^{s}$, Dec =$-58^{o}07^{\prime}24^{\prime\prime}$,
while its Galactic coordinates are l= 299$^{o}$.313, b=4$^{o}$.556. It is therefore located very close to Trumpler~20 in longitude,
but significantly higher onto the Galactic plane.\\

\noindent
As a part of their southern sky search for star clusters, Moffat \& Vogt (1973) describe NGC~4337 with the following words:
{\it  In spatial order, the stars 9, 6, 10, 7, 11, 13, 14, 2,4,5 and 8 appear to form an open arc, revealing some similarity to the "stellar rings" of Issestedt(1968). However the photometry shows  no physical connection among these stars; they most likely represent a random arrangement of field stars. The actual cluster may be a small group (diameter $\sim 2^{\prime}$) of stars located between nos. 14 and 15, which are too faint to be reached by the 61 cm telescope except for the possible members 12, 13, 14, 1 and 15, the photometry of these also shows no physical association.}

To help the reader, Moffat \& Vogt (1973) map  is  reproduced in the left panel of Fig~1, while the right panel shows the color-magnitude diagram (CMD ) one can derive from their stars. This CMD gives full justice to their preliminary conclusions. Clearly no hints are visible of  the presence of a  physical group,  at odds with the photographic plate
which gives the clear impression of  some star concentration. 

This latter evidence motivated us to acquire new CCD data. We briefly describe the observations
in Section~2 of this paper, and use the photometric data-set to probe the cluster structure in Section~3.
Section~4 is devoted to the analysis of the CMDs of NGC~4337. Some conclusions are given in Sections~5.

\begin{figure}
\includegraphics[width=\columnwidth]{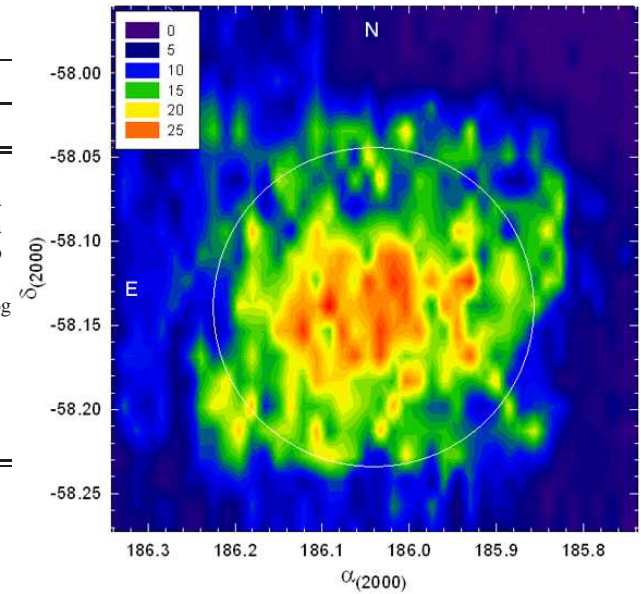}
\caption{A density contour map in the area of NGC~4337. The map is 20 arcmin on a side, North is up, East to the left. The white circle indicates the estimate cluster area, the radius being 6 arcmin.}
\end{figure}

\section{Observations}
We took UBVI images of NGC~4337 in a 20$\times$20 squared arcmin area during two observing runs, on  2010 Feb 14  and 2010 Mar 11, at Cerro Tololo Inter-American Observatory, using the 0.9m and the 1.0m telescope. 
The first night  (at the 0.9m) was not photometric, and 
therefore we tied all the images to the second night (at the 1.0m), which was photometric. During this second night, we took
multiple images of the standard star fields PG~1047 and SA98, covering an airmass range 1.15$-$2.2. Five frames per filter, with exposures ranging from  20 to 2400 secs  (U),  5 to 1800 secs (B), 5 to 1200 secs (V and I) were obtained, to avoid bright star saturation and at the same time to get deep enough to unravel the cluster main sequence. 
These observations were part of a long term project aimed at collecting
high-quality data for a large open cluster data-set. We refer the reader to previous publications (e.g. Carraro \& Costa 2007, and Carraro \& Seleznev 2012)
for all the details of data reduction and photometric calibration. The final catalog, cross-correlated with 2MASS, contains 
8198 entries, and is made available at the CDS database. An excerpt of the catalog is presented in Table~2, which lists
the clump star candidates.

\section{Cluster structure and radius}
In Fig.~2 we show a contour plot in the area of NGC~4337, derived from our photometry. Star counts have been performed using a grid, and  density inside each field computed via a suitable kernel estimate (see V\'azquez et al. 2010 for more details).
The cluster has quite an irregular shape, but certainly appears as a prominent over-density. 
The density peak ($\sim$ 65 stars/arcmin$^{2}$) is offset with respect to the center of the field, which was chosen according to the cluster nominal center. A visual inspection of Fig~2 provides us with an estimate
of the cluster radius of $\sim$  7 arcmin (see the super-imposed white circle ), 
larger than visual inspection on the DSS (Dias et al. 2002).\\ 
Our quantitative star counts  analysis fully  supports  Moffat \& Vogt (1973) visual impression that
NGC~4337 is clearly a star over-density with respect to the surrounding field.

\begin{figure}
\includegraphics[width=\columnwidth]{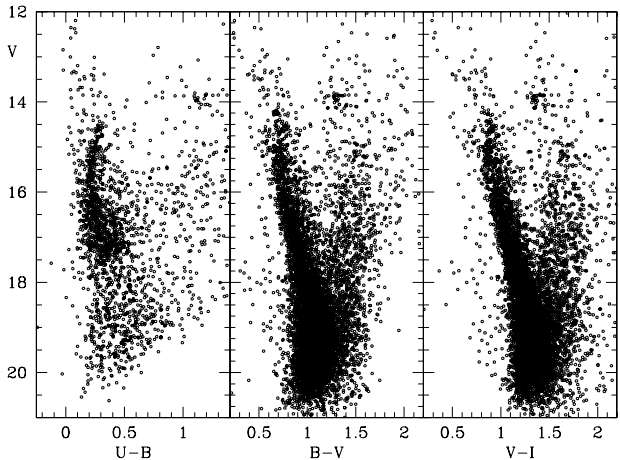}
\caption{CMDs for all the stars observed in the field of NGC~4337.}
\end{figure}

\section{Analysis of the photometric diagrams}
The CMD for all the stars in the observed field is shown in Fig.~3.  This strikingly  resembles the CMD of an intermediate-age/old
open cluster.  
Although field star contamination is significant, a turn off point (TO) can be easily identified at V $\sim$ 15.50,  together
with a prominent clump of stars at V $\sim$ 14.0, U-B $\sim$ 1.10, B-V $\sim$ 1.30 , and V-I $\sim $ 1.35.
This diagram, in tandem with start counts,  confirms beyond any reasonable doubt that NGC~4337 is indeed a physical star cluster.\\

\noindent
To alleviate field star contamination we make use of the previous section results, and consider in the following only stars within
5 arcmin from the cluster center. The resulting  CMD in the V/B-V plane is presented in Fig.~4.
The cluster main features emerge very neatly, and should immediately remind the reader the CMD of the more famous old open clusters IC~4651
and NGC~3680 (see Table~1),  or NGC~752.  The TO is located at V = 15.50, and the MS extends down to V = 19, in spite of some residual field star contamination. 
A scattered sequence of binary stars is visible to the right of the MS.  The binary sequence intersects the MS at V $\sim$15.1. We cannot exclude the presence of a few blue straggler stars.\\
Interestingly enough, the MS TO region has a curved shape, typical 
of intermediate-age/old clusters.  Finally,
the Hertzsprung gap  is quite evident, together with the red giant  clump, made of a dozen stars. \\

\noindent
We stress also the fact that NGC~4337 CMD is much cleaner than Trumpler~20 CMD (Seleznev et al. 2010).\\

\noindent
To infer for the first time the cluster basic parameters,
we super-imposed on the cluster CMD the ridge line of IC~4651 (red dashed line), taken from Anthony-Twarog et al. (1988) study.   
IC~4651 seems o be the best choice, being a classical old open clusters, with an age around 1.7 Gyr (Anthony-Twarog et al. 1998, Meibon 2000, Anthony-Twarog et al. 2009), and a metallicity
[Fe/H] =+ 0.11 (Carretta et al. 2004). This is slightly super-solar, as expected for clusters in the inner disc.\\

The fit has been performed by shifting IC~4651 ridge line by  2.6 mag in magnitude and 0.14 mag in color.
The fit is really impressive
as far as the MS is concerned. The only difference is in the magnitude of clump, since IC~4651 mean clump magnitude is $\sim$ 0.3
mag. brighter than NGC~4337.
Since the fit is very convincing, we conclude that the two clusters most probably share the same metallicity, but NGC~4337 would be somewhat younger than IC~4651.
IC~4651 reddening and apparent  distance modulus are 0.12 and 10.40 mag, respectively, according to the recent study by Anthony-Twarog et al. (2009). 
This implies that NGC~4337 has a reddening E(B-V) $\sim$  0.26 and an apparent distance modulus of $\sim$13.00. From these figures we derive
a preliminary cluster helio-centric  distance of 2.8 kpc, which, in turn, implies a distance of $\sim$ 7.5 kpc from the Galactic center.

\begin{figure}
\includegraphics[width=\columnwidth]{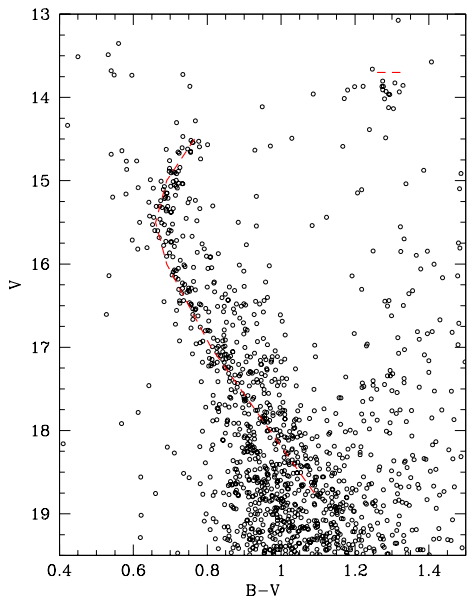}
\caption{CMDs for all the stars  within NGC~4337 radius. The red line is IC~4651 ridge line offset by 0.14 in color and 2.6 in magnitude}
\end{figure}

\begin{table*}
\caption{An excerpt of the photometric data presented in this paper. The full table is posted at the CDS archive. The table lists all the clump star candidates}
\begin{tabular}{lcccccccccc}
\hline
ID & RA(2000.0 & Dec(2000.0) & V & $\sigma_{V}$ & (B-V) & $\sigma_{(B-V)}$ & (U-B) & $\sigma_{(U-B)}$ & (V-I) & $\sigma_{(V-I)}$\\
\hline
& [deg] & [deg] & mag & mag & mag & mag & mag & mag & mag & mag\\
\hline\hline
     90& 186.06567&  -58.14727&  13.804&   0.002&   1.275&   0.003&   1.101&   0.005&   1.311&   0.004\\
     91& 186.16358&  -58.13047&  13.827&   0.002&   1.308&   0.002&   1.199&   0.006&   1.404&   0.003\\
     99& 186.16451&  -58.22229&  13.853&   0.002&   1.304&   0.004&   1.254&   0.010&   1.406&   0.004\\
   100& 185.86345&  -58.04729&  13.861&   0.009&   1.347&   0.020&   9.999&   9.999&   1.449&   0.024\\
   106& 186.01314& - 58.11106&  13.867&   0.002&   1.277&   0.002&   1.007&   0.005&   1.348&   0.004\\
   107& 186.02449&  -58.12087&  13.868&   0.002&   1.273&   0.003&   1.095&   0.005&   1.336&   0.004\\
   110& 185.98837&  -58.12068&  13.908&   0.002&   1.275&   0.003&   1.127&   0.004&   1.322&   0.004\\
   114& 186.03714&  -58.11592&  13.928&   0.002&   1.284&   0.002&   1.101&   0.005&   1.336&   0.004\\
   115& 185.98609&  -58.12367&  13.933&   0.002&   1.320&   0.003&   1.180&   0.005&   1.365&   0.005\\
   117& 185.96857&  -58.11566&  13.961&   0.002&   1.291&   0.003&   1.123&   0.006&   1.332&   0.003\\
   118& 185.98385&  -58.11050&  13.966&   0.002&   1.294&   0.003&   1.115&   0.005&   1.353&   0.004\\
   121& 186.03666&  -58.03773&  13.977&   0.002&   1.311&   0.003&   1.101&   0.010&   1.378&   0.004\\
   124& 186.13244&  -58.13392&  14.017&   0.001&   1.280&   0.002&   1.135&   0.006&   1.321&   0.003\\
   128& 186.18727&  -58.03610&  14.056&   0.004&   1.332&   0.006&   1.155&   0.014&   1.436&   0.006\\
   130& 186.02041&  -58.08607&  14.061&   0.002&   1.330&   0.003 &  1.190&   0.008&   1.365&   0.004\\
 \hline\hline
\end{tabular}
\end{table*}

\section{Summary and conclusions}
In this work we have revisited the star cluster NGC~4337, previously considered as a random arrangement
of field stars.  One of the  main feature of the cluster CMD is an extremely well defined MS TO region, resembling closely
IC~4651, NGC~752, and NGC 3680, all extremely well-known textbook intermediate-age old open clusters.
NGC~4337 harbors a dozen clump stars, which would make it as massive as IC~4651, and most probably in the same dynamical stage (Meibom 2000).
In fact, we do not see any manifestation of low mass stars depletion,  as for NGC~3680 and NGC~752.\\

\noindent
The cluster is therefore  a rare example of an old open cluster in the inner disc. According to our preliminary estimates of the cluster parameters, 
its Galacto-centric distance would make NGC~4337 one of the more distant cluster from the Sun among the well known old open clusters located inside
the solar ring.
This is a privileged location to establish in a more solid way the slope of the inner disc radial abundance gradient (Magrini et al. 2010), and its evolution with time.
We expect this study to prompt a spectroscopic campaign to derive its metal abundance, and better constrain its fundamental parameters.

\section*{Acknowledgments}
We thank the Centre de Donn\'{e}es Astronomiques de Strasbourg (CDS), the
U. S. Naval Observatory and NASA for the use of their electronic facilities,
especially SIMBAD, ViZier and ADS, and the WEBDA database.  GC thanks Nikolaus Vogt for granting the
permission to reproduce his original plot  in the left panel of Fig~1 of this paper.
EC acknowledges support by the Fondo Nacional de Investigaci—n
Cient'fica y Tecnol—gica (project No. 1110100 Fondecyt), and the Chilean
Centro de Excelencia en Astrofisica y Tecnolog'as Afines (PFB06). We finally thank the referee, Bruce Twarog,
for his comments.



\end{document}